\documentclass[final,5p,times,twocolumn]{elsarticle} 
\usepackage{lineno,hyperref}
\modulolinenumbers[5]

\journal{Journal of Nuclear Instruments and Methods in Physics Research A}

\usepackage{url}
\begin{document}

\begin{frontmatter}

\title{Neutron spectroscopy: The case of the spherical proportional counter}

\author[aff1]{I.~Giomataris}
\author[aff2]{I.~Katsioulas}
\author[aff2]{P.~Knights}
\author[aff2]{I.~Manthos\corref{cor1}}
\ead{i.manthos@bham.ac.uk}
\author[aff2]{T.~Neep}
\author[aff2]{K.~Nikolopoulos}
\author[aff1]{T.~Papaevangelou}
\author[aff2]{R.~Ward}

\address[aff1]{IRFU, CEA, Universite Paris-Saclay, F-91191 Gif-sur-Yvette, France}
\address[aff2]{School of Physics and Astronomy, University of Birmingham, B15 2TT, United Kingdom}

\cortext[cor1]{Corresponding author}

\begin{abstract}
Neutron spectroscopy is an invaluable tool for many scientific and industrial applications, including underground Dark Matter searches.  Neutron-induced backgrounds produced by cosmic ray muons and the cavern radioactivity can mimic the expected Dark Matter signal. However, existing neutron detection methods have several drawbacks and limitations, thus measurements remain elusive.  A promising new approach to neutron spectroscopy is the use of a nitrogen-filled spherical proportional counter that exploits the $^{14}$N(n,$\alpha$)$^{11}$B and $^{14}$N(n, p)$^{14}$C reactions.  This is a safe, inexpensive, effective and reliable technique. In this work, the latest instrumentation developments are incorporated in a compact detector operated at the University of Birmingham (UoB) with high gain at gas pressure up to 1.8\,bar.  We demonstrate spectroscopic measurements of thermalised and fast neutrons respectively from an $^{241}$Am-$^9$Be source and from the MC40 cyclotron facility at UoB. Additionally, the detector response to neutrons is simulated using a framework developed at UoB and compared with the experimental results.

\end{abstract}


\end{frontmatter}

%\linenumbers

\section{Introduction}
Efficient neutron spectroscopy is a highly desirable tool with applications including science and industry \citep{applications}. In particular, in underground Dark Matter (DM) experiments the neutron-induced background from the remaining cosmic ray flux and the cavern radioactivity can mimic the expected DM signal, thus limiting the experimental sensitivity. Despite the extensive efforts, neutron detectors are complex and results are scarce, while the measurement of the neutron energy distributions remain typically an insurmountable challenge. Additionally, the unique characteristics of each underground cavern are difficult to be experimental precisely determined, limiting the accuracy of Monte Carlo studies. A precise, in situ, spectroscopic measurement of the neutron background would be of great benefit for rare-event experiments, providing all the information to efficiently characterise and mitigate this source of background. 

The most commonly used method relies on the $^3$He(n,p)$^3$H reaction, but the large target mass required to avoid the recoil reaching the detector's wall (the so-called ``wall effect'') along with the scarcity of $^3$He undermine this method. On the other hand, existing alternative methods present several disadvantages \citep{Kouzes:2015tsc}. A novel approach is the use of the nitrogen-filled spherical proportional counter (SPC) \citep{spc} that relies  on the $^{14}$N(n,p)$^{14}$C and $^{14}$N(n,$\alpha$)$^{11}$B processes, exhibiting cross-sections comparable to those of the $^3$He(n,p)$^3$H reaction for thermal and fast neutrons respectively. A proof of principle has already been demonstrated \citep{neutron} but limitations in the SPC instrumentation at the time prevented the full exploitation of the method's potential. Recent developments in detector instrumentation, namely the multi-anode sensor \citep{achinos,achinos2}, confront these limitations and makes the operation of the detector efficient on high gain and pressure conditions \citep{neutron_paper}. 

In this work, we demonstrate spectroscopic neutron measurements with a 30\,cm diameter SPC, responding to neutrons from an $^{241}$Am$^9$Be source and from the MC40 cyclotron facility at the University of Birmingham (UoB) \citep{cyclotron}. Results from a dedicated simulation framework \citep{simulation} are compared to the experimental data.
\section{The spherical proportional counter and the neutron detection method}
\begin{figure}[h]
\centering
\includegraphics[width=0.83\linewidth]{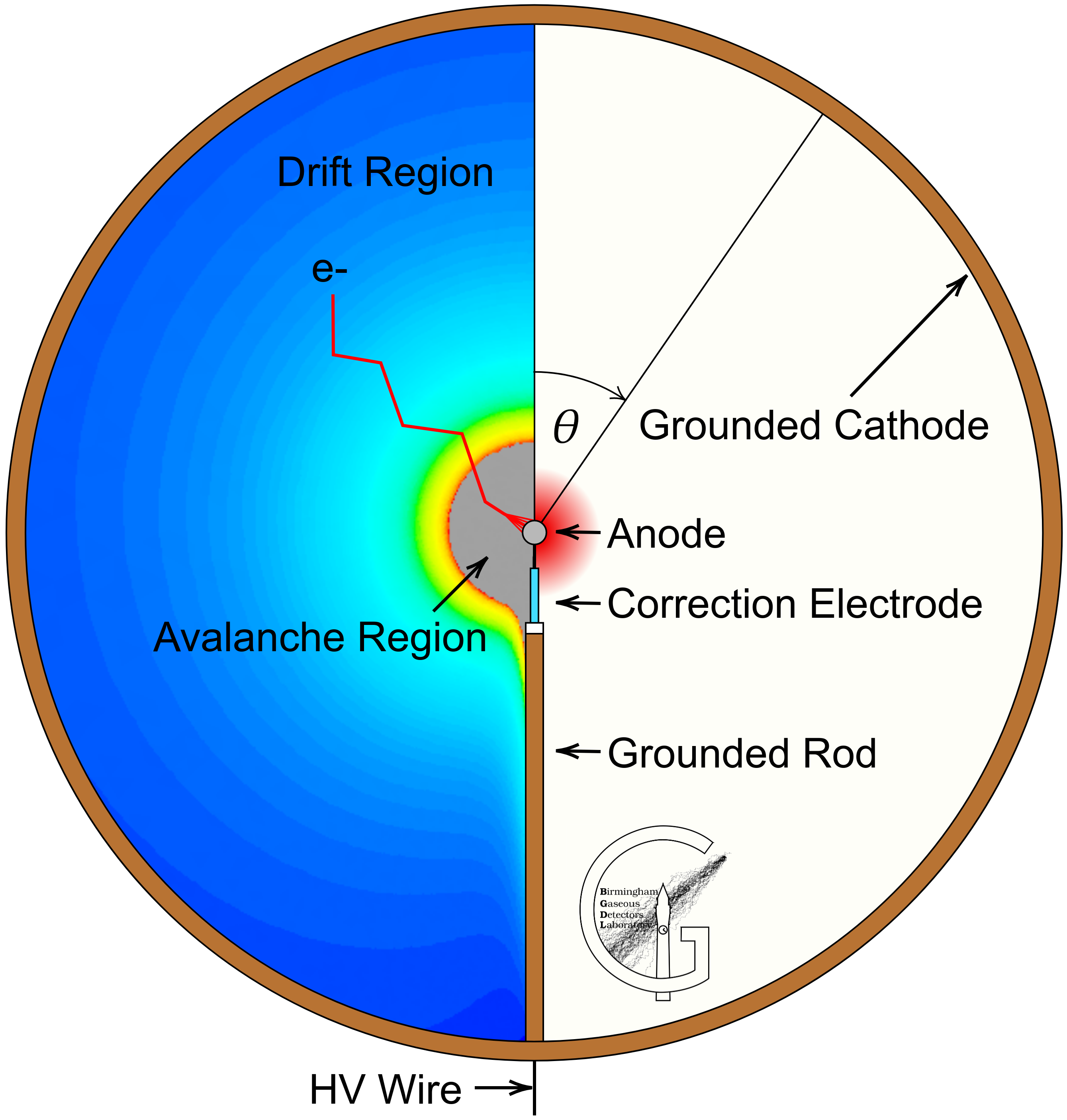}
\caption{Schematic of the spherical proportional counter.\label{fig:fig1}}
\end{figure}
The SPC is a gaseous detector that finds several applications \citep{spc_applications}. It is based on the spherical geometry, as shown in Fig.~\ref{fig:fig1}, with the grounded vessel serving as a cathode, while the anode is supported by a metallic grounded rod through which the high voltage is applied. The main characteristic of the SPC is that the cathode radius, $r_{c}$, is much larger than the anode radius, $r_{a}$. This results in the electric field being approximately radial with magnitude $E\approx \frac{V}{r^2} r_{a}$, where V is the applied voltage and $r$ is the distance from the detector centre. As a result, the volume of the detector is divided into a drift and an amplification region. Additionally, an important asset of the SPC, again thanks to $r_{c}\gg r_{a}$, is that the capacitance scales as $C\approx 4\pi\epsilon\epsilon_{0}r_{a}$. This is independent of the detector size (i.e. the cathode size) resulting in low electronic noise and making the SPC ideal for rare event searches.

\begin{figure}[h]
\centering
\includegraphics[width=0.9\linewidth]{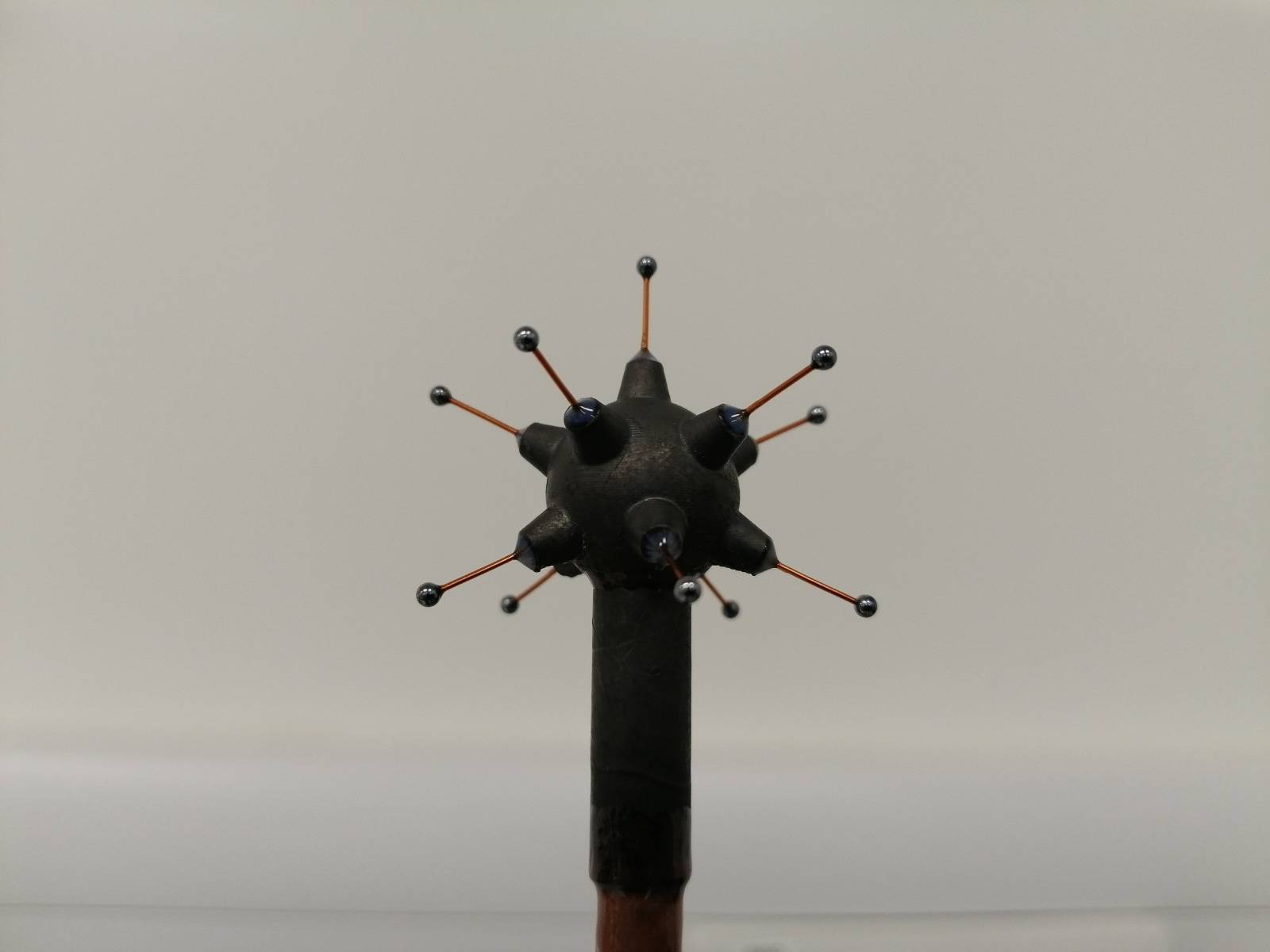}
\caption{The 11-anode ACHINOS sensor.\label{fig:fig2}}
\end{figure}

As shown in Fig.~\ref{fig:fig1}, the simplest form of the SPC has a single anode at its centre. However, the detector operation is governed by the ratio of the electric field magnitude to the gas pressure $E/P$, and a single anode limits the electric field magnitude that can be achieved without compromising detector stability. Thus, nitrogen-filled SPCs operating with a single anode are limited in their operating pressure by the electric field magnitude.  The ACHINOS multi-anode sensor \cite{achinos2,achinos}, a recent development in SPC instrumentation, overcomes this challenge. Multiple small radius anodes (that determine the amplification field) are placed at a  constant distance from the detector centre in a regular pattern, forming a virtual sphere of larger radius (that determines the drift field). In this way, the drift and amplification fields are decoupled and can be tuned independently. Fig. \ref{fig:fig2} presents the 11-anode sensor used in the present measurements. The possibility to fiducialise the detector is provided by reading out the anodes in two groups:  one that includes the 6 farthermost (``far'' side) anodes to the rod, and another comprising the 5 nearest (``near'' side).

The method exploits the 
\begin{eqnarray*}
 ^{14}\textrm{N} + \textrm{n} \rightarrow {}^{14}\textrm{C} + \textrm{p} + 625\,{\textrm{keV}}\\
 ^{14}\textrm{N} + \textrm{n} \rightarrow {}^{11}\textrm{B} + \alpha - 159\,{\textrm{keV}}
\end{eqnarray*}
reactions through the detection of the Townsend avalanche  initiated by the charged product of each reaction in the gas of the SPC . The first reaction is exothermic, releasing 625\,keV, and has sufficient cross section to allow thermal neutron detection. The second reaction presents comparable cross section to the $^3$H for fast neutron energies up to 20\,MeV. In addition,  the increased target mass due to the higher atomic number of N$_2$ compared to $^3$He, suppresses the wall effect. For these reasons, this method is a safe, cost-efficient and reliable alternative for spectroscopic neutron measurements. 
\section{Neutron detection at the graphite stack}
\begin{figure}[h]
\centering
\includegraphics[width=0.9\linewidth]{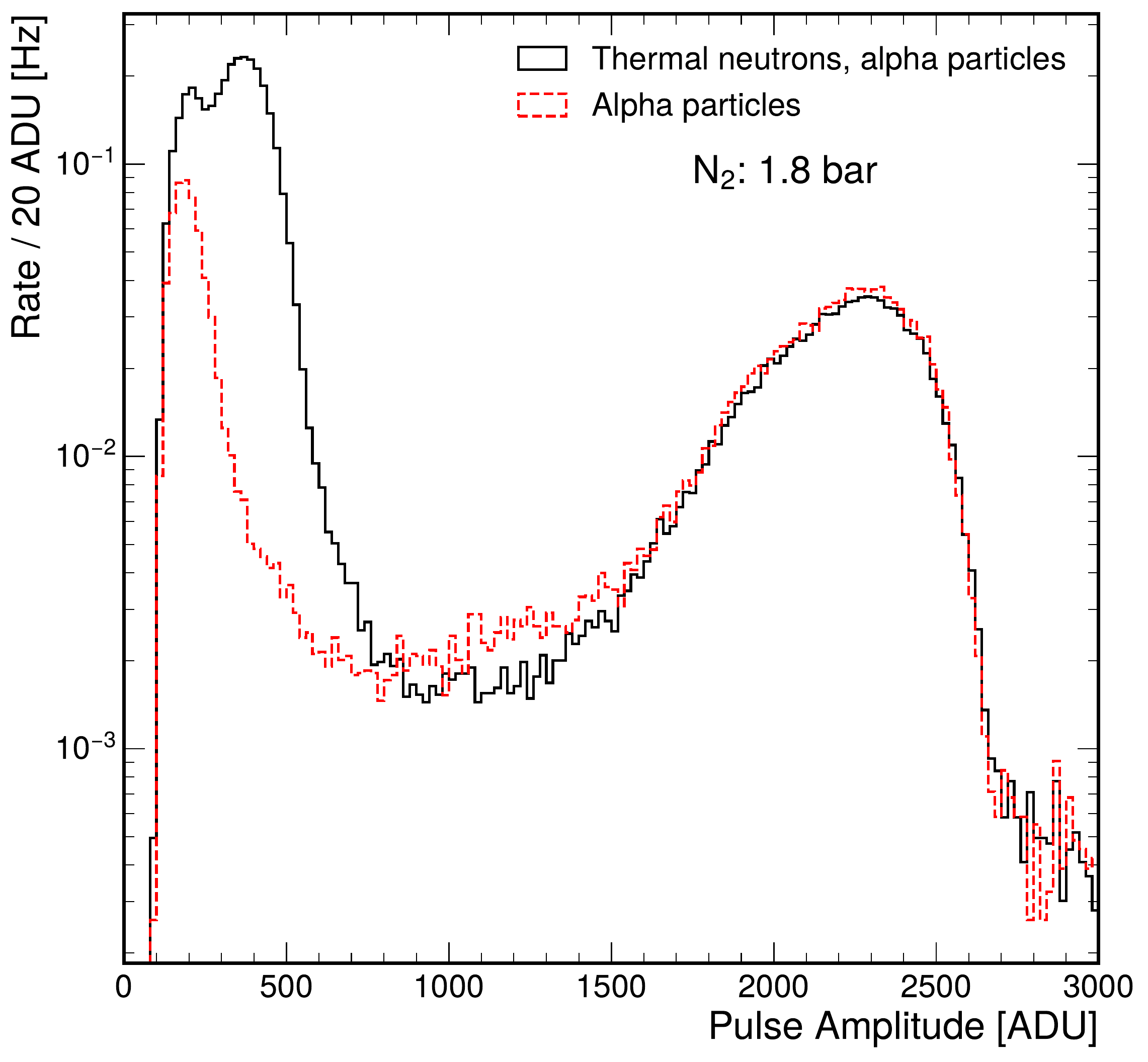}
\caption{Pulse amplitude distribution of the response of the SPC on $^{241}$Am-$^9$Be neutrons thermalised by the graphite stack and on 5.30\,MeV $\alpha$-particles from a $^{210}$Po source. With dashed line the response of the SPC only to the $\alpha$-particles from the $^{210}$Po source is given, \cite{neutron_paper}. \label{fig:fig3}}
\end{figure}
The detection of neutrons with this method is demonstrated using thermalised neutrons from an $^{241}$Am$^9$Be source inside a graphite stack at UoB. Details of the detection setup and the results of the performed campaign are reported in \cite{neutron_paper}. In this work, we highlight the results with the detector operating with N$_2$ at 1.8\,bar pressure and 6\,kV applied at the anode, which constitutes record conditions for neutron spectroscopy with the SPC. 
The solid line in Fig. \ref{fig:fig3} represents the pulse amplitude distribution produced on the far side of the detector in response to thermalised neutrons (expected at 625\,keV) and to 5.30\,MeV $\alpha$-particles emitted from a $^{210}$Po source. The $^{210}$Po source is placed at the inner side of the cathode, opposite the rod, providing a direct calibration. In absence of the neutron source, the far side of the detector responds only to $\alpha$-particles as shown by the dashed line in Fig. \ref{fig:fig3}. Standard pulse shape parameters (i.e. the rise time and the pulse width) are used to discriminate signal from noise.
\section{Fast neutron measurements at the MC40 cyclotron}
\begin{figure}[h]
\centering
\includegraphics[width=0.9\linewidth]{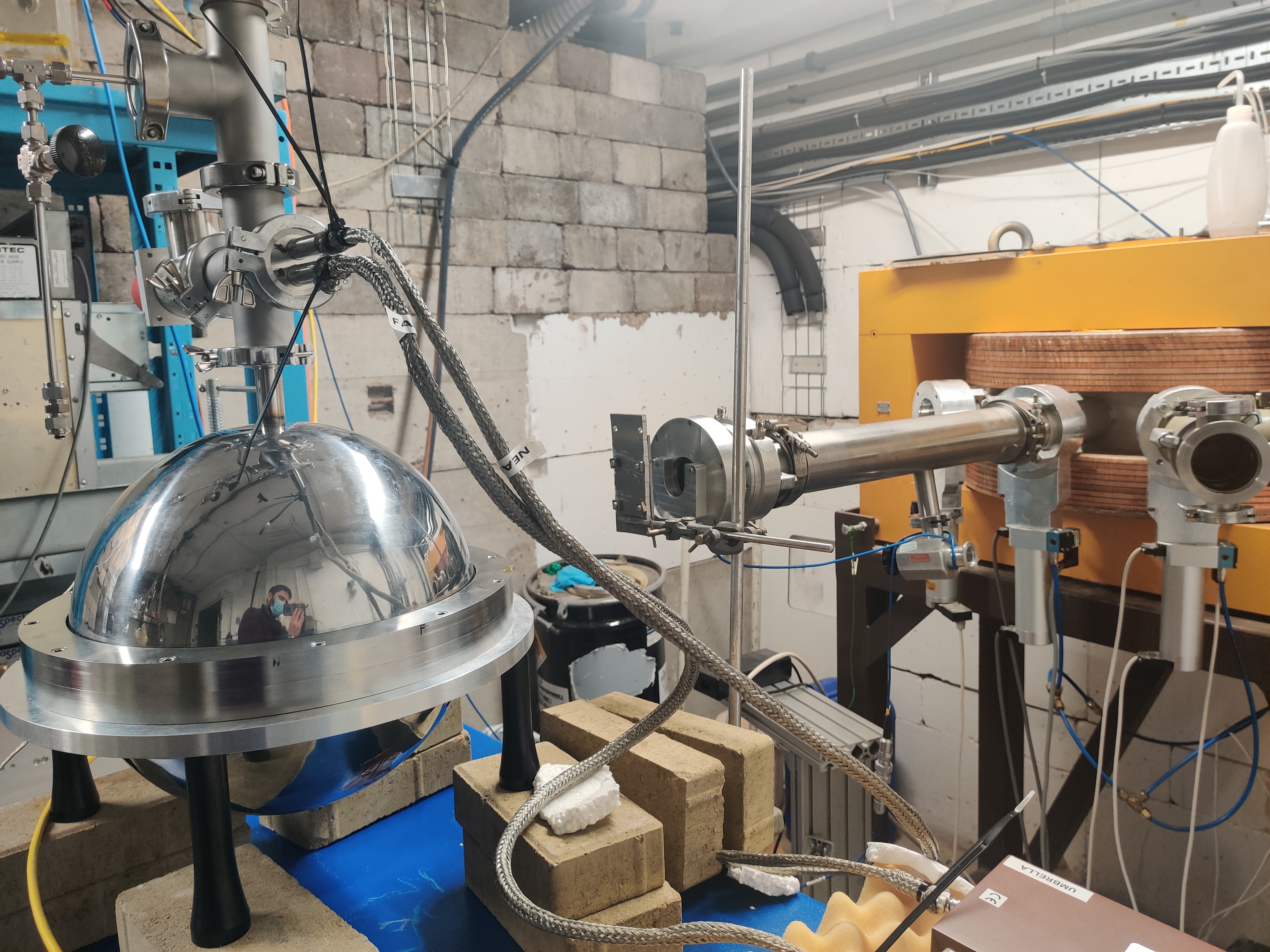}
\caption{The experimental setup at the MC40 cyclotron  facility. 5.90\,MeV deuterons are hitting the $^9$Be target, producing neutrons with energies up to 8\,MeV and then detected by the SPC.  \label{fig:fig4}}
\end{figure}
\begin{figure}[h]
\centering
\includegraphics[width=0.9\linewidth]{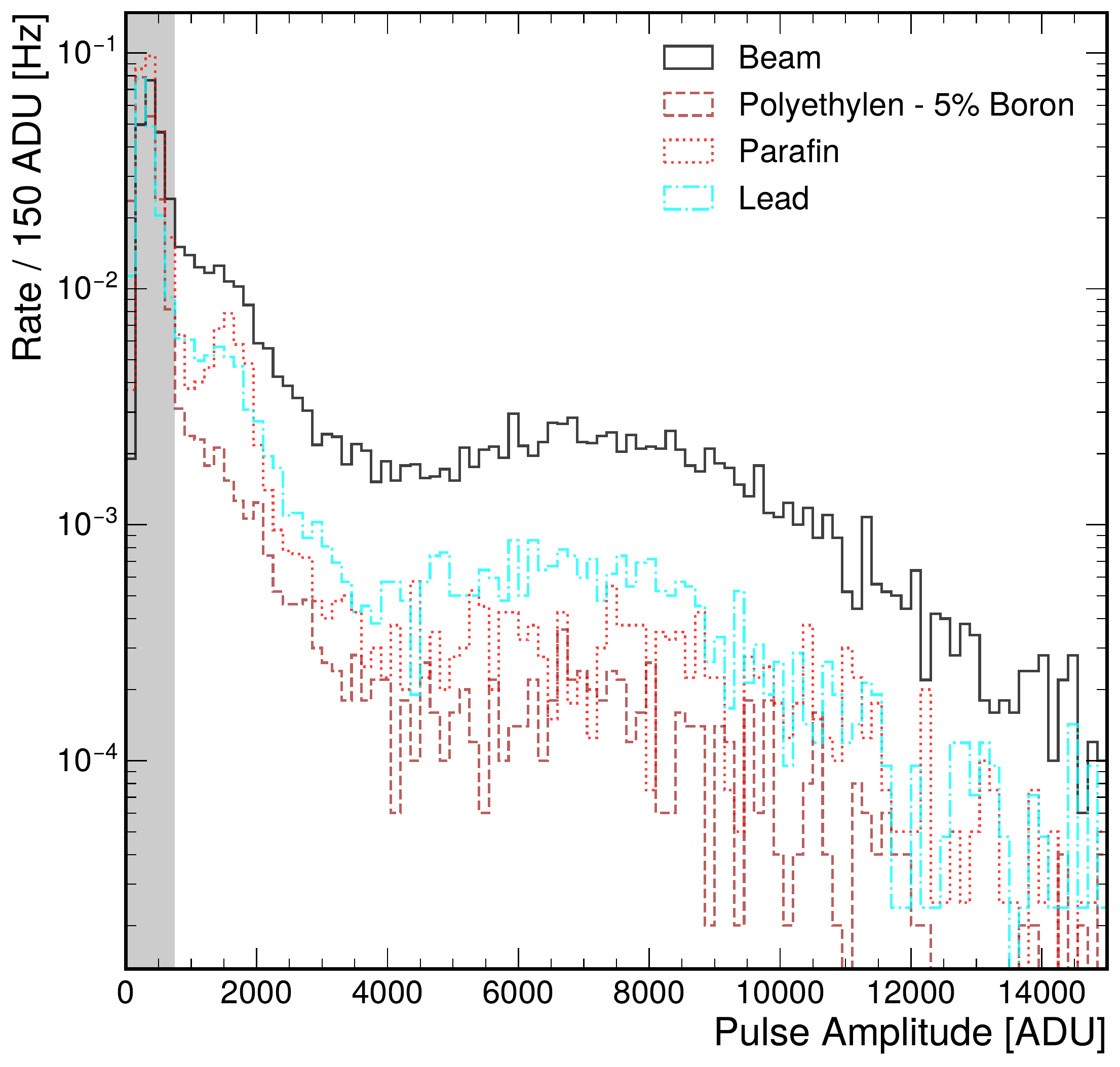}
\caption{Pulse amplitude distribution of the detector response on direct neutron beam (line) and moderated with: polyethylene doped with 5\% $^{10}$B (dashed), paraffin (dotted) and lead (dash/dotted). The shadowed area is noise dominated.
\label{fig:fig5}}
\end{figure}

Neutron detection is also demonstrated with the response of the detector to neutrons from the UoB MC40 cyclotron facility. 5.90$\pm$0.08\,MeV deuterons from the cyclotron are incident on a 3\,mm thick $^9$Be target, as shown in Fig. \ref{fig:fig4}, and produce neutrons through the $^9$Be(d,n)$^{10}$B reaction with energy up to 8\,MeV \citep{berillium}. The detector setup is the same as the one described in Section 3, and the same pulse shape parameters (i.e. rise time and pulse width) were used for event selection. The detector energy calibration is provided with 5.30\,MeV $\alpha$-particles from a $^{210}$Po source. Spectroscopic measurements of neutrons have been performed with the SPC filled with 1\,bar nitrogen and operating with 4.4\,kV anode voltage.  In order to verify the neutron detection, materials providing neutron energy moderation and neutron absorption were used. In addition, pressure and voltage settings were selected with the objective of high gain operation with a dynamic range adequate to accommodate the full neutron energy spectrum provided by the cyclotron. In the pulse amplitude distribution of Fig. \ref{fig:fig5} the solid line represents the detection of neutrons without any moderation, while the dotted line represents the response of the SPC using 16\,cm of paraffin as a moderator, between the $^9$Be target and the SPC. This results in a significant drop of the fast neutron rate and a corresponding  distinguishable thermal neutron peak at $\sim$1500\,ADU, with a pulse amplitude that is in agreement with the calibration results.
The dashed line represents the response of the detector when Boron doped polyethylene (5\% w/w) was used as a moderator. The effect was not only to thermalise fast neutrons, as can be seen by the large drop of the rate, but to suppress the thermal neutrons. $^{10}$B that has a significant cross section of 3840\,barns for thermal neutron absorption. Finally, despite not being a neutron moderator, the effect of lead on the neutron beam was tested and shown with the dashed/dotted line. This resulted only to a rate drop -due to neutron scattering in the lead- but with the preservation of the exact shape on the amplitude distribution that implies no energy moderation. Through background measurements it was ascertained that the amplitude up to $\sim$750\,ADU was dominated by the cyclotron background and is not taken into consideration.
\begin{figure}[h]
\centering
\includegraphics[width=0.9\linewidth]{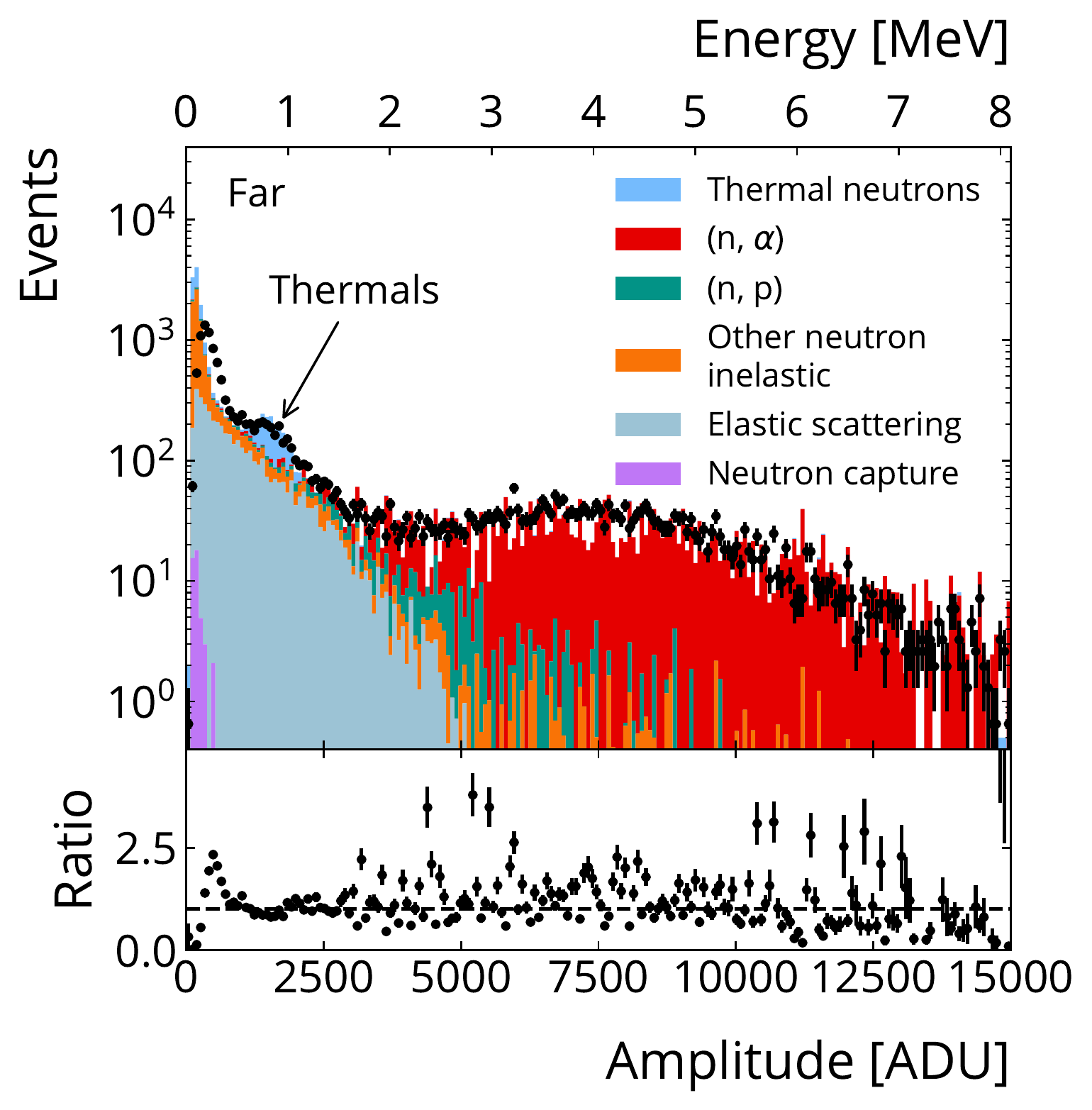}
\caption{The responce of the SPC on the neutrons from the MC40 cyclotron in comparison with the simulation results that indicate the origin of the detected events. \label{fig:fig6}}
\end{figure}
 
The experimental response of the SPC on neutrons emitted from the $^9$Be target is compared with the simulation in Fig. \ref{fig:fig6}. The SPC simulation framework developed at the UoB was used, \citep{simulation}, which combines the strength of ANSYS \citep{ansys}, Garfield++ \citep{garfield} and GEANT4 \citep{geant4} simulation packages. It simulates the interaction of the detector with particles, signal formation including the electronics response and provides the detector response at the data acquisition level. In addition, for the specific experimental apparatus, the simulation included the interaction of deuterons with the $^9$Be target.
In Fig. \ref{fig:fig6}, the simulated SPC response to neutrons emitted from the $^9$Be target (histograms) is compared with the experimental data (points). The simulation provides insights into the interaction that initiated each event. The interactions are presented with different colours. %Additionally, the simulation provides the  energy equivalence of ADUs, that within the energy resolution of the detector confirms both the thermal neutrons peak at 625\,keV and the energy range of the detected neutrons. 
The dominance of the $^{14}$N(n, $\alpha$)$^{11}$B reaction for fast neutrons (above 2\,MeV) is apparent, while the $^{14}$N(n, p)$^{14}$C reaction has significant contribution up to $\approx$3\,MeV. As already reported in Ref. \cite{tipp}, simulation studies proved that the discrimination of pulses originated by these two reactions is possible using pulse shape parameters. The agreement between data and simulation is remarkable, with an emphasis on the thermal neutrons. Moreover, a non negligible contribution is attributed to elastic scatterings of the neutrons with the target gas. 
\section{Summary}
The spherical proportional counter provides an efficient, cost-effective alternative for neutron spectroscopic measurements that can serve several applications, including neutron background characterisation in underground Dark Matter experiments. In this work we demonstrated the successful operation of the spherical proportional counter to neutron spectroscopy with operating pressure up to 1.8\,bar with neutron thermalised by the UoB graphite stack infrastructure, and high gain operation of the SPC detecting neutrons with energy up to 8\,MeV from the MC40 cyclotron facility at the UoB.

\section*{Acknowledgments}

This project has received funding from the European Union's Horizon 2020 research and innovation programme under the Marie Sk\l{}odowska-Curie grant agreements No 845168 (neutronSphere), No 841261 (DarkSphere) and No 101026519 (GaGARin).

\bibliography{elba_proceedings}

\end{document}